\documentclass[prb,twocolumn,amsmath]{revtex4}% Physical Review B

\usepackage{graphicx}% Include figure files
\usepackage{dcolumn}% Align table columns on decimal point
\usepackage{bm}% bold math
\usepackage[T1]{fontenc}

\newcommand{\az}{a$_0$}
\newcommand{\acrit}{a$_{c}$}

\begin{document}

%\newpage
%$~~~~$
%\vspace{5cm}

\title{Finite-size effects in BaTiO$_3$ nanowires}
\author{Gr\'egory Geneste$^1$, Eric Bousquet$^1$, Javier Junquera$^2$ and Philippe
Ghosez$^1$}
\address{$^1$ D\'epartement de Physique, Universit\'e de Li\`ege, B\^atiment B-5, B-4000 Sart Tilman, Belgium \\
$^2$ CITIMAC, Universidad de Cantabria, Avda. de los Castros s/n, E-39005, Santander, Spain}
%\date{\today}

\begin{abstract}
The size dependence of the ferroelectric properties of BaTiO$_3$ nanowires
is studied from first-principles. We show that the ferroelectric distortion along
the wire axis disappears below a critical diameter of about 1.2 nm. This
disappearance is related to a global contraction of the unit cell resulting from
low atomic coordinations at the wire surface. It is shown that a ferroelectric
distortion can be recovered under appropriate tensile strain conditions.
\end{abstract}
\maketitle

% INTRODUCTION
Ferroelectric oxides constitute an important class of
multifunctional compounds, attractive for numerous technological
applications \cite{lines}. Pushed by the miniaturization constraints
imposed by future devices, there is now an increasing interest in
the synthesis and use of nanosized ferroelectrics \cite{science1}.
Nowadays, various ferroelectric nanostructures -- ultrathin
films \cite{tybell99,science2}, nanowires \cite{jjurban-wsyun},
nanotubes \cite{nanoshell,morrison}, nanoislands \cite{disloc} -- can be
grown with control at the atomic scale. However, bringing
ferroelectrics to the nanoscale raises fundamental questions.

Ferroelectricity is a collective effect, resulting from a delicate balance
between short-range and long-range Coulomb interactions.
In confined structures, both interactions are modified with respect
to the bulk and it is commonly believed that ferroelectricity
is altered and eventually totally suppressed when the system
reaches a critical size sometimes referred to as
the ``ferroelectric correlation volume'' \cite{lines}.
First-principles simulations are a fundamental tool to understand
size effects in ferroelectrics~\cite{GhosezU}. Recent first-principles studies
provided valuable insight into the behavior of
thin films and highlighted the crucial role of the depolarizing
field\cite{junquera2003,kornev,Dawber05}. 
A few works based on effective Hamiltonians\cite{Fubellaiche,naumov} and
shell models\cite{stachiotti} also considered other low dimensional
systems, such as nanorods and nanodisks. However, despite these efforts,
very little is known on the evolution of ferroelectric
properties in nanowires with respect to their size and shape.

In this letter, we investigate from first-principles the influence of finite-size
effects on the ferroelectric properties of BaTiO$_3$ nanowires. We
identify a critical diameter below which the ferroelectric distortion along the
wire axis is totally suppressed. We assign this behavior to surface atomic
relaxation effects. Below the critical diameter, a ferroelectric
state can be recovered under appropriate tensile strain conditions.
The crucial role of the longitudinal interatomic coupling along
Ti--O chains and the weak interchain interactions are also discussed.

%%%%%%%%%%%%%%%%%%%%%%%%%%%%%%%%%%%%%%%%%%%%%%%%%%%%%%%%%%%
% PART 1 : nanowires
%%%%%%%%%%%%%%%%%%%%%%%%%%%%%%%%%%%%%%%%%%%%%%%%%%%%%%%%%%%

\begin{figure}[htbp] {\par\centering
 {\scalebox{0.35}{\includegraphics{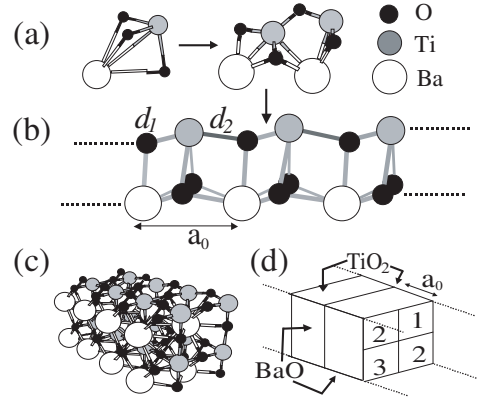}}} \par}
 \caption{(Color online) BaTiO$_3$ stoichiometric wires: BaTiO$_3$ clusters (a)
assemble into infinite chains (b), themselves gathered into nanowires (c). The
diameter of the nanowires is defined from the number of Ti--O chains ($n$)
assembled together within the wire. Panels (c) and (d) correspond to $n=4$.}
\label{figure1}
\end{figure}

We focus on stoichiometric infinite BaTiO$_3$ nanowires.
As illustrated in Fig.~\ref{figure1}, we consider model systems built
from individual BaTiO$_3$ clusters that are assembled into infinite
chains, themselves combined into wires of increasing diameters. These
stoichiometric wires have two BaO and two TiO$_2$ lateral surfaces,
as displayed in Fig.~\ref{figure1}(d).
This asymmetry is responsible for a dipole perpendicular
to the wire axis. However, it will not be further discussed since it is not relevant to
the present study~\cite{Note-1}. In what follows, we will
focus on the eventual disappearance of the ferroelectric instability {\it along}
the wire direction $z$.
The equilibrium lattice parameter
along $z$ will be referred to as $a_0$,
and the wire diameter will be specified
by an integer $n$ that corresponds to the number of Ti-O chains that are
gathered together within the wire. The alternating Ti-O distances
along $z$ will be called $d_1$ and $d_2$, as shown in Fig.~\ref{figure1}(b).

Our calculations have been performed within the local density approximation (LDA)
to density functional theory (DFT) using a numerical atomic orbital method
as implemented in the {\sc Siesta} code\cite{siesta}. Technical
details are similar to those of Ref.~\onlinecite{junquera2003_prb}. Some of the calculations have been double checked using the {\sc abinit} package~\cite{abinit}.
Periodic replicas of the wires are separated by more than 10 \AA\ of vacuum.
We explicitly checked that the replica of the wire generated by periodic boundary
conditions is not affecting our results~\cite{Note-2}.
The atomic positions have been relaxed until the maximum component of the
force on any atom is smaller than 0.01 eV/\AA.

\begin{table}
\caption{\label{distances} {Interatomic distances and equilibrium lattice
parameter $a_0$ along the wire axis, for the $n=1$ wire. O$_I$ (resp. O$_{II}$)
refers to oxygen atoms along the Ti--O (resp. Ba--O) chains. Results for the bulk
cubic paraelectric unit cell of BaTiO$_3$ are shown for comparison. Units
in {\AA}.}  }
\begin{ruledtabular}
\begin{tabular}{ccccc} & Ti-O$_I$ & Ba-O$_{II}$ & Ba-Ti & $a_0$ \\
%   &  ({\AA}) & ({\AA}) & ({\AA}) & ({\AA}) \\
\hline
Wire & 1.88 & 2.52 & 3.07 & 3.60 \\
Bulk & 1.97 & 2.78 & 3.41 & 3.94 \\
\end{tabular}
\end{ruledtabular}
\end{table}

For the thinnest wire ($n=1$, Fig.~\ref{figure1}(b))
the bond lengths and the equilibrium lattice constant $a_0$ of
the relaxed structure reported in Table~\ref{distances} are significantly shorter than in bulk. This
can be explained from the low coordination of the different atoms
in the structure, the lack of some interactions being compensated
by the strengthening of the the remaining bonds.
The system remains insulating with a bandgap of 3.3 eV.
The relaxed structure presents a plane of symmetry perpendicular to $z$
($d_1=d_2$) and is therefore not spontaneously polarized along the wire
direction, demonstrating that ferroelectricity has disappeared. However,
imposing a tensile strain, we can recover a polar structural distortion along
the Ti-O chains ($d_1 \neq d_2$) above a critical lattice constant
$a_c = 4.05$ {\AA}. This strain induced phase transition is second order.
Above $a_c$, both $d_1-d_2$ and the polarization,
calculated from the Berry phase approach,
\cite{kingsmith,sanchezportal} increase linearly with $a_0$.

\begin{figure}[htbp] {\par\centering 
{\scalebox{0.3}{\includegraphics{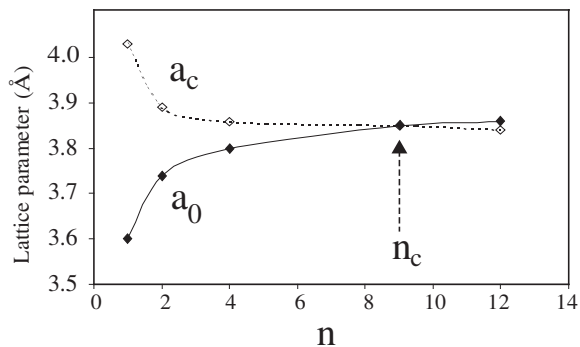}}} \par}
 \caption{Evolution of {\acrit} (white diamonds) and {\az} (black diamonds) as a function of the number $n$ of Ti--O chains
in a transverse section of the wire.}
\label{figure2}
\end{figure}

The fully relaxed $n=4$ wire, shown in Fig.~\ref{figure1}(c),
is also non-polar along $z$. However, $a_c$ has been reduced to 3.88 {\AA}
and $a_0$ is now larger (3.80 {\AA}) as expected from the increase of the
mean atomic coordination numbers with respect to the $n=1$ wire.
Figure~\ref{figure2} represents the evolution of $a_0$ and $a_c$ for wires
of increasing $n$. We observe that the difference, $a_c-a_0$, progressively
decreases. A crossover between both curves is observed at $n_c = 9$, that
corresponds to a diameter of 1.2 nm. Above $n_c$, unconstrained wires are
ferroelectric whereas below this value, a spontaneous polarization only appears
under tensile strain.

In Fig.~\ref{figure2}, $a_c$ keeps a nearly constant value around 3.85 {\AA}
(except for $n=1$) while the disappearance of the ferroelectric distortion
at small sizes originates from the progressive decrease of $a_0$.
The ferroelectric behavior of nanowires
appears therefore to be monitored by the strong sensitivity of ferroelectricity
to the unit cell volume, the latter affected by low coordination
effects at free surfaces. As $n$ increases, we recover a polar
distortion, however, its amplitude only progressively increases
as $a_0$ slowly evolves to its bulk value.

\begin{figure}[htbp] {\par\centering
{\scalebox{0.58}{\includegraphics{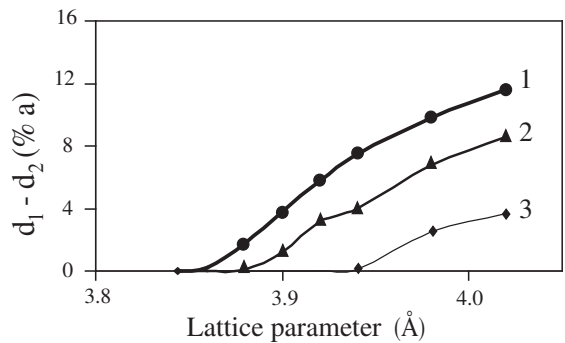}}} \par}
 \caption{Local distortions as a function of the lattice parameter for the
$n=4$ wire. Labels of the chains as in Fig.~\ref{figure1}(d).}
\label{figure3}
\end{figure}

Similarly, below $n_c$, a second-order phase transition to a polar state
is observed under tensile strain, but the resulting polarization strongly
depends on the lattice parameter. This is illustrated in Fig.~\ref{figure3} for $n=4$,
where we quantify the polar distortion through the
anisotropy of the interatomic distances ($d_2-d_1$) along the
distinct Ti-O chains. This provides a qualitative measure of the
contributions to the polarization of the individual chains since
the Born effective charges (along $z$) are reasonably similar from
one chain to the other (variations smaller than 20 \% \cite{zti}).
The strong inhomogeneity of the distortions for the different Ti-O chains
highlights a deep influence of the surface: the ferroelectric
distortion parallel to the wire surface is enhanced at
TiO$_2$ surfaces and reduced around the BaO surfaces,
a feature similar to what was previously reported for free-standing
slabs \cite{padilla}.
Note that the transition to the polar state
does not occur at the same lattice constant in the different chains,
giving rise, within a given range of lattice constants, to mixed
ferroelectric-paraelectric states in these nanosized systems.

The distinct behavior of the different Ti-O chains within the same
wire suggests weak inter-chain interactions. This is further confirmed
by the fact that the polarization of each chain can be reversed almost
independently, with very small changes in the total energy
of the system and in the amplitude of the polar displacements.
It is the longitudinal coupling of the atomic displacements inside individual
Ti--O chains that appears as the main origin of the ferroelectric
distortion. This behavior is similar to what was previously reported for the
bulk \cite{ghosez98}. In the phonon dispersion curves of BaTiO$_3$\cite{ghosez99},
the unstable ferroelectric mode at $\Gamma$ remains similarly unstable
at $X$ and $M$ but {\it not} at $R$, pointing out a strong
anisotropy of the interatomic force constants \cite{ghosez98}, a significant
coupling of the atomic displacement along Ti--O chains, and the absence of strong
interchain interactions. The resulting "chain-like" character of the ferroelectric
instability is intrinsic to BaTiO$_3$. It is preserved in nanowires and provides a general argument to understand the survival of ferroelectricity along the wire axis down
to very small radii. The situation might be different in other compounds such as
PbTiO$_3$ in which the ferroelectric instability is more isotropic~\cite{ghosez99}.

%%%%%%%%%%%%%%%%%%%%%%%%%%%%%%%%%%%%%%%%%%%%%%%%%%%%%%%%%%%%%%%%%%%
%   CONCLUSION
%%%%%%%%%%%%%%%%%%%%%%%%%%%%%%%%%%%%%%%%%%%%%%%%%%%%%%%%%%%%%%%%%%%

In summary, the ferroelectric behavior of infinite stoichiometric BaTiO$_3$
nanowires has been studied from first-principles. In BaTiO$_3$, the ferroelectric
instability exhibits a marked "chain-like" character so that, in nanowires, it
could a priori be preserved down to very small sizes. Nevertheless, at small
radii, low atomic coordinations at the surface produce a contraction of the unit
cell that is responsible for the suppression of the ferroelectric distortion at a
critical radius estimated around 1.2 nm. Below this radius the ferroelectric
distortion has disappeared at the equilibrium volume but can be recovered under
appropriate tensile strain conditions.

\section*{Acknowledgements}
The authors thank M. Dawber for careful reading of the manuscript.
This work was supported by the VolkswagenStiftung (I/77 737), FNRS-Belgium (2.4562.03), the
R\'egion Wallonne (NOMADE) and the European Network of Excellence "FAME".


\begin{thebibliography}{10}
\bibitem{Note-1}
Similar calculations were performed for selected non-stoichiometric wires with full Ba-O and
full TiO$_2$ coverage, yielding results similar to Fig.~\ref{figure2} except that the
crossing between $a_c$ and $a_0$ curves was slightly shifted to larger (resp. smaller)
cell parameters for BaO (resp. TiO$_2$) coverage.
\bibitem{Note-2}
To test the possible influence (especially in reason of the lateral dipole) of the
artificial replica induced by the periodic boundary conditions, we repeated some calculations
enlarging the vacuum regions up to 20 \AA, without significant change in the results.
\bibitem{zti}The Born effective charges along $z$ for the four
Ti in a section of the wire of Fig.~\ref{figure1}(c) are: 
$Z^*_{Ti1} = 5.10$, $Z^*_{Ti2} = 5.78$, $Z^*_{Ti3} = 5.93$ (labels as in Fig.~\ref{figure1}(d)).
\end{thebibliography}
\end{document}